\documentclass[showpacs,preprintnumbers,amsmath,amssymb,eqsecnum]{revtex4}
\usepackage{graphicx}
\begin{document}
\preprint{IFF-RCA-08-02}
\title{Quantum cosmic models and thermodynamics}

\author{Pedro F. Gonz\'{a}lez-D\'{\i}az}
\email{p.gonzalezdiaz@imaff.cfmac.csic.es} \affiliation{Colina de
los Chopos, Centro de F\'{\i}sica ``Miguel A. Catal\'{a}n'',
Instituto de F\'{\i}sica Fundamental,\\ Consejo Superior de
Investigaciones Cient\'{\i}ficas, Serrano 121, 28006 Madrid
(SPAIN).}

\author{Alberto Rozas-Fern\'{a}ndez} \email{a.rozas@cfmac.csic.es}
\affiliation{Colina de los Chopos, Centro de F\'{\i}sica ``Miguel
A.
Catal\'{a}n'', Instituto de F\'{\i}sica Fundamental,\\
Consejo Superior de Investigaciones Cient\'{\i}ficas, Serrano 121,
28006 Madrid (SPAIN).}

\date{\today}
\begin{abstract}
The current accelerating phase of the evolution of the universe is
considered by constructing most economical cosmic models that use
just general relativity and some dominating quantum effects
associated with the probabilistic description of quantum physics.
Two of such models are explicitly analyzed. They are based on the
existence of a sub-quantum potential and correspond to a
generalization of the spatially flat exponential model of de
Sitter space. The thermodynamics of these two cosmic solutions is
discussed, using the second principle as a guide to choose which
among the two is more feasible. The paper also discusses the
relativistic physics on which the models are based, their
holographic description, some implications from the classical
energy conditions, and an interpretation of dark energy in terms
of the entangled energy of the universe.
\end{abstract}

\pacs{95.36.+x, 98.80.-k}

\maketitle

\noindent Keywords: Quantum potential, tachyonic dark energy,
Lorentz invariance, Cosmic phantom models.

\pagebreak

\section{Introduction}

The last years have seen an influx of ideas and developments
conceived to try to find a cosmic model able to consistently
predict the observational data that each time more clearly imply
that the current universe is accelerating (see [1] for a recent
review). Nevertheless, none of such models can be shown to
simultaneously satisfy the following two requirements, (i) exactly
predicting what observational data point out, and (ii) an economic
principle according to which one should not include unnecessary
ingredients such as mysterious cosmic fluids or fields nor
modifications of the very well tested background theories such as
general relativity. The use of scalar fields in quintessence or
k-essence scenarios is not with standing quite similar to
including an inflaton in inflationary theories for the early
universe [2]. Even though, owing to the success of the
inflationary paradigm which actually shares its main
characteristics with those of the present universal acceleration,
many could take this similarity to be a reason enough to justify
the presence of a scalar field also pervading the current
universe, it could well be that a cosmic Occam's Razor principle
would turn out to be over and above the nice coincidence between
predictions of usual models for inflation and what has been found
in cosmic observations such as the measurement of background
anisotropies. After all, the medieval opinion that the simplest
explanation must be the correct explanation has proved to be
extremely fruitful so far and, on the other hand, the paradigm of
inflation by itself still raises some deep criticisms. Occam's
Razor is also against the idea of modifying gravity by adding to
the relativistic Lagrangian some convenient extra terms.

Besides general relativity, quantum theory is the other building
block which can never be ignored while constructing a predicting
model for any physical system. Although it is true that a quantum
behavior must in general be expected to manifest for small-size
systems, cosmology is providing us with situations where the
opposite really holds. In fact, fashionable phantom models for the
current universe are all characterized by an energy density which
increases with time, making in this way the curvature larger as
the size of the universe becomes greater. In such models quantum
effects should be expected to more clearly manifest at the latest
times where the universe becomes largest. Thus, it appears that
quantum theory should necessarily be another ingredient in our
task to build up an economical theory of current cosmology without
contravening the Occam's Razor philosophy.

A cosmological model satisfying all the above requirements has
been recently advanced [3]. It was in fact constructed using just
a gravitational Hilbert-Einstein action without any extra terms
and taking into account the probabilistic quantum effects on the
trajectories of the particles but not the dynamical properties of
any cosmic field such as quintessence or k-essence. The resulting
most interesting cosmic model describes an accelerating universe
with an expansion rate that goes beyond that of the de Sitter
universe into the phantom regime where the tracked parameter of
the universal state equation becomes slightly less than -1, and
the future is free from any singularity. Such a model, although
still a toy one, will thus describe what can be dubbed a {\it
benigner} phantom universe because, besides being regular along
its entire evolution, it does not show the violent instabilities
driven by a non-canonical scalar-field kinetic term as by
construction the model does not have a negative kinetic term nor
it classically violates the dominant energy condition which
guarantees the stability of the theory, contrary to what the
customary phantom models do. Another cosmic model was also
obtained which describes an initially accelerating universe with
equation of state parameter always greater than -1, that
eventually becomes decelerating for a while, to finally contract
down to a vanishing size asymptotically at infinity. The latter
model seems to be less adjustable to current observational data
although we are not completely sure as this is a toy model.

We know very little about the theoretical nature and origin of
dark energy. Therefore, it is worth exploring its thermodynamic
properties seeking a deeper understanding, in the hope that this
consideration will shed some light on the properties of dark
energy and help us understand its rather elusive nature. Actually,
some attention has been paid to the subject of thermodynamics of
dark energy when this is interpreted as a radiation field [4] and
a phantom field [5]. Other authors have also studied a variety of
dark energy properties related to thermodynamics [6-10]. Besides
reviewing the essentials of it, in this paper we are going to deal
with two fundamental aspects of the benigner phantom scenario. On
the one hand, we shall investigate in some detail the basic
physics on which it is grounded, and on the other hand, we shall
consider some thermodynamical aspects of the benigner phantom
scenario putting special emphasis on general functions such as
entropy, enthalpy as well as temperature, and study the implied
holographic description, some consequences from the quantum
violation of the classical conditions on energy, and finally the
interpretation of the models in terms of the entanglement energy
of the accelerating universe.

The paper can be outlined as follows. In Sec. II we briefly review
the cosmic quantum models, and in Sec. III we discuss the
thermodynamics that can be associated with such models and its
implications  in the violation of the classical energy conditions,
the cosmic holography, and their connection to the notion of
entanglement entropy for an accelerating universe. We conclude and
add some further comments in Sec. IV. An Appendix is added where
new material is presented on the consistency of the cosmic quantum
models and the quantum aspects that we must include in the theory
of special relativity on which such models are based.

\section{The quantum cosmic models}
In this section we briefly review the basic ideas and formulae of
the cosmic quantum models which were considered in Ref. [3] (For a
previous work from which the ideas provided in [3] were derived ,
see Ref. [11].) These models are a quantum extension from the
known tachyon dark energy model [12,13]. The latter scenario is
physically grounded on the relativistic Lagrangian for a particle
of mass $m_0$, i.e. $L=-m_0\sqrt{1-v^2}$ (with $v=\dot{q}$ the
particle velocity), up-grading the coordinate $q$ to a scalar
field $\phi$, the squared velocity to
$\partial_i\phi\partial^i\phi\equiv\dot{\phi}^2$, and the rest
mass to the scalar field potential $V(\phi)$. In order to
introduce the cosmic quantum models, we first derive the
Lagrangian that corresponds to a particle which is subject to the
usual quantum effects. Thus, we apply the Klein-Gordon equation to
a general quasi-classical wave function
$\Psi=R(q,t)\exp(iS(q,t)/\hbar)$, and obtain from the resulting
real part the expression for the momentum
\begin{equation}
p=\sqrt{E^2 +\tilde{V}_{SQ}^2 -m_0} ,
\end{equation}
where $E$ is the classical energy and
$\tilde{V}_{SQ}=\hbar\sqrt{(\nabla^2 R-\ddot{R})/R}$ is the
sub-quantum potential, so that the Lagrangian becomes
\begin{equation}
\tilde{L}=\int d\dot{q}p= -m_0
E\left(\arcsin\sqrt{1-v^2},\sqrt{1-\frac{\tilde{V}_{SQ}^2}{m_0^2}}\right),
\end{equation}
in which $E(x,k)$ is the elliptic integral of the second kind.
Following Bagla et all [12] we upgrade then the quantities
entering Lagrangian (2.2) to scalar field quantities in such a way
that $\dot{q}^2=v^2\rightarrow\partial_i \phi\partial^i \phi\equiv
\dot{\phi}^2$, and $m_0\rightarrow\tilde{V}(\phi)$, with
$\tilde{V}(\phi)$ the scalar field potential, and hence we obtain
$\tilde{L}=-\tilde{V}(\phi)E(x(\phi),k(\phi))$, where
$x(\phi)=\arcsin\sqrt{1-\dot{\phi}^2}$ and
$k(\phi)=\sqrt{1-\tilde{V}_{SQ}^2/\tilde{V}(\phi)^2}$. Now, it was
shown in Ref. [3] that for the model to imply an accelerating
universe characterized by an energy density and pressure which
depend both on the sub-quantum potential only and vanish (when no
cosmological constant is present) in the limit $\hbar\rightarrow
0$, the above Lagrangian must be expressed as a Lagrangian density
to read [3]
\begin{equation}
L=-V(\phi)\left[E(x,k)-\sqrt{1-\dot{\phi}^2}\right] ,
\end{equation}
where we have subtracted the tachyonic Lagrangian density derived
from classical special relativity and $k$ can be written as
$k=\sqrt{1-V_{SQ}^2/V(\phi)^2}$, with $V_{SQ}=\tilde{V}_{SQ}/a^3$
and $V(\phi)=\tilde{V}(\phi)/a^3$ the respective sub-quantum and
scalar field potential energy densities, $a$ being the scale
factor of the universe. Lagrangian density (2.3) in fact vanishes
in the limit $\hbar\rightarrow 0$ and from it one can derive the
pressure, $p$, and energy density, $\rho$
\begin{equation}
p=L
\end{equation}
\begin{equation}
\rho=V(\phi)\left(\frac{\sqrt{\dot{\phi}^2+\frac{V_{SQ}^2}{V(\phi)^2}(1-
\dot{\phi}^2)}\dot{\phi}}{\sqrt{1-\dot{\phi}^2}}
+E(x,k)-\frac{1}{\sqrt{1-\dot{\phi}^2}}\right) .
\end{equation}
Letting the equation of state parameter $w=p/\rho$ to be
time-dependent and using the general expression [12,13]
$\dot{\rho}/\rho=-3H(1+w)=2\dot{H}/H$, with $H=\dot{a}/a$, one can
obtain [3]
\begin{equation}
\rho=6\pi G\left(\dot{H}^{-1}H\dot{\phi}V_{SQ}\right)^{2}
\end{equation}
\begin{equation}
p=w(t)\rho=-\left(1+\frac{2\dot{H}}{3H^2}\right)\rho
\end{equation}
with
\begin{equation}
\dot{H}=\pm 4\pi G\dot{\phi}V_{SQ},\;\;\; H=H_0 \pm4\pi G\phi
V_{SQ}t .
\end{equation}
Regularity requirements for $\ddot{\phi}$ on the equation of
motion derived from the Lagrangian density (2.3) leads, by
manipulating [3] the Friedmann equations and the above equations,
to the condition $\dot{\phi}^2 =1$ and to the simpler expressions
\begin{equation}
\rho=6\pi G\left(\dot{H}^{-1}HV_{SQ}\right)^{2}
\end{equation}
\begin{equation}
p=w(t)\rho=-\left(1+\frac{2\dot{H}}{3H^2}\right)\rho ,
\end{equation}
where
\begin{equation}
\dot{H}=\pm 4\pi GV_{SQ},\;\;\; H=H_0 \pm4\pi GV_{SQ}t ,
\end{equation}
so erasing all traces of the scalar field $\phi$. What remains
instead are some constants and a time-dependence which vanishes
when $\hbar\rightarrow 0$; that is, if we disregarded the
integration constant $H_0$ (which plays the role of a cosmological
constant) only purely quantum effects are left. It is worth
remarking that we do not expect the sub-quantum potential
$\tilde{V}_{SQ}$ appearing in Eq. (2.2) to remain constant along
the universal expansion but to increase like the volume $V=a^3$ of
the universe does, with $a$ the scale factor. It is the
sub-quantum potential density $V_{SQ}=\tilde{V}_{SQ}/V$ appearing
in (2.3) what should be expected to remain constant at all cosmic
times.
\begin{figure}
\includegraphics[width=.9\columnwidth]{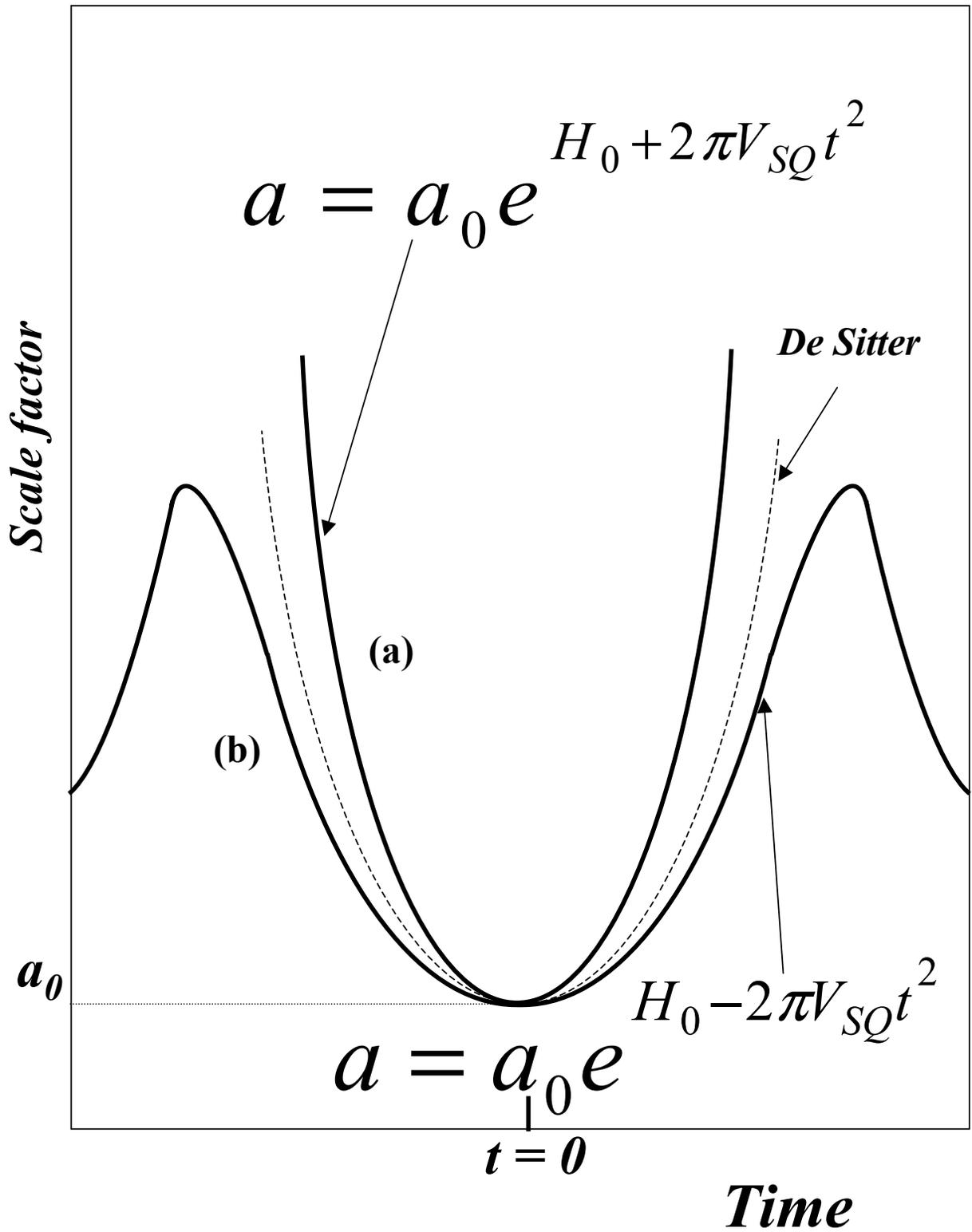}

\caption{\label{fig:epsart} Cosmic solutions that result from the
introduction of a sub-quantum potential density $V_{SQ}$ when
$\dot{\phi}^2=1$. Solution (a) goes like in de Sitter space with
the same $H_0$, but with higher acceleration. Solution (b)
corresponds to the case where $H_0^2>4\pi V_{SQ}$ and represents a
universe which is initially expanding in an accelerated way (at a
rate slower than in de Sitter space with the same $H_0$), then
expands in a decelerated way for a while to finally contract
toward a zero radius as $t\rightarrow\infty$. On the figure we
have used units such that $\hbar=c=G=1$}
\end{figure}

Integrating finally Eq. (2.11) we obtain for the scale factor of
the universe
\begin{equation}
a(t)=a_0 e^{H_0 t \pm 2\pi GV_{SQ}t^2} ,
\end{equation}
with $a_0$ the initial value $a_0=a(0)$. From the set of solutions
implied by Eq. (2.12), we shall disregard from the onset the one
corresponding to $H_0=0$ and $t=\sqrt{\frac{\ln(a_0/a_{-})}{2\pi
GV_{SQ}}}$ (which corresponds to the sign - for the term
containing the sub-quantum potential) as it would predict the
unphysical case of a universe which necessarily is currently
contracting. The chosen solutions are depicted in Fig.1 as
compared to the usual de Sitter solution. Both of such solutions
become flat de Sitter in the classical limit $\hbar\rightarrow 0$.
Besides, we should mention that the de Sitter limit is not exactly
identical to the observable universe. By putting matter-energy
momentum in the theory, we would then expect different features
from the ones found in the de Sitter universe, such as
instability, or that the exclusion limit could vary with matter
inside, at least during the current epoch of mixture of matter and
dark energy. Some $w>-1$ branches could survive or some $w<-1$
branches could be excluded. One can thus draw the conclusion that
pure quantum probability effects on the particles filling the
universe make by themselves the universe to accelerate quicker or
slower than what is predicted by a cosmological constant, but do
not induce a future big rip singularity in any case. In the next
section we shall see that it is the phantom regime ($w<-1$)
predicted by the solution with the + sign what agrees with the
thermodynamic second law and gives therefore rise to what we can
name a benigner phantom regime that is free from singularities or
unphysical negative kinetic terms in the Lagrangian. We should
expect that the inclusion of a very little proportion of matter
would not change the above conclusion. In Sec. IV it will be also
shown that these quantum effects can be interpreted as a cosmic
entanglement energy.

The reader who may be interested in a discussion on further
aspects that re-enforce the consistency of the models considered
above and on the quantum modifications that such a description
entails in the background relativistic theory is addressed to the
Appendix.

\section{benigner phantom thermodynamics}
\subsection{Thermodynamics}

The thermodynamical description of dark energy has offered an
alternative route to investigate the evolution of the current
universe [5-10]. However, whereas well-defined expressions can be
obtained for dark energy models with equations of state $p=w\rho$
where $w>-1$, in the phantom regime characterized by $w<-1$ either
the temperature or the entropy must be definite negative. In what
follows we shall discuss the thermodynamical properties of the
benigner cosmic models in which it will be seen that these
problems are alleviated. By using the above equations we proceed
now to derive expressions for the thermodynamical functions
according to the distinct models implied by the sign ambiguity in
Eq. (2.12) and the possibility that the cosmological term be zero
or not, only for the solution branches that correspond to a
positive time $t>0$. On the one hand, the translational energy
that can be associated with the scalar field would be proportional
to [5] $a^3\dot{\phi}^2$ and therefore, because $\dot{\phi}^2 =1$
[3], the essentially quantum temperature associated with the
sub-quantum models must be generally given by
\begin{equation}
T_{SQ}=\kappa a^3 ,
\end{equation}
with $\kappa$ a given positive constant whose value will be
determined later. It is worth noting that, unlike for phantom
energy models [14], in this case the temperature is definite
positive even though the value of the state equation parameter $w$
be less than -1. Moreover, this temperature is an increasing
function of the scale factor and hence it will generally increase
with time. It must be also stressed that $T_{SQ}$ must be a
quantum temperature as it comes solely from the existence of a
sub-quantum potential.

On the other hand, one can define the entropy and the enthalpy.
If, since the universe evolves along an irreversible way,
following the general thermodynamic description for dark energy
[4,5], one defines the total entropy of the sub-quantum medium as
$S_{SQ}(a)= \rho {\rm V}/T_{SQ}$, with ${\rm V}=a^3$ the volume of
the universe, then in the case that we choose for the scale factor
the simplest expanding solution (without cosmological constant)
$a_{+}=a_0\exp(2\pi GV_{SQ}t^2)$, with $V_{SQ}$ the sub-quantum
potential density, we obtain the increasing, positive quantity
\begin{equation}
S_{SQ}(a_{+})=\frac{V_{SQ}}{\kappa}\ln\left[\left(\frac{a_{+}}{a_0}\right)^3\right]
.
\end{equation}
This definition of entropy satisfies the second law of
thermodynamics.

For the kind of systems we are dealing with one may always define
a quantity which can be interpreted as the total enthalpy of the
universe by using the same expression as for entropy, but referred
to the internal energy which, in the present case, is given by
$\rho +p$, instead of just $\rho$. Thus, we can write for the
enthalpy $H_{SQ}=(\rho+p){\rm V}/T_{SQ}$. which leads for the same
cosmic solution to the constant, negative definite quantity
\begin{equation}
H_{SQ}(a_{+})=-\frac{V_{SQ}}{\kappa} ,
\end{equation}
whose negative sign actually implies a quantum violation of the
dominant energy condition and indicates that we are in the phantom
regime.

The consistency of the above definitions of entropy and enthalpy
will be guaranteed  in what follows because the expressions that
we obtain from them in the limit $V_{SQ}\rightarrow 0$ are the
same as for de Sitter space.

Since the third power of the ratio $a_{+}/a_0$ must be
proportional to the number of states in the whole universe, the
mathematical expression of the entropy given by Eq. (3.2) could
still be interpreted to be just the statistical classical
Boltzmann's formula, provided we take the constant $V_{SQ}/\kappa$
to play the role of the Boltzmann's constant $k_{B}$, or in other
words, $k_B$ is taken to be given by $k_B=V_{SQ}/\kappa$, in such
a way that the temperature becomes $T_{SQ}(a_{+})=V_{SQ}a^3/k_B$
which consistently vanishes at the classical limit
$\hbar\rightarrow 0$. If we let $\hbar\rightarrow 0$ then it would
be $T_{SQ}(a_{+})$ but not $S_{SQ}(a_{+})$ what vanishes. In this
way, Eq. (3.3) becomes
\begin{equation}
H_{SQ}(a_{+})=-k_B .
\end{equation}
The negative value of this enthalpy can be at first sight taken as
a proof of an unphysical character. However, one could also
interpret $H_{SQ}(a_{+})$ the way Schr\"{o}dinger did [15] with
the so-called "negentropy" as a measure of the information
available in the given system, which in the present case is the
universe itself.

The above results correspond to the case in which the universe is
endowed with a vanishing cosmological constant. If we allow now a
nonzero cosmological term $H_0$ to exist, i.e. if we first choose
the solution $a_{-}=a_0\exp(H_0 t-2\pi GV_{SQ}t^2)$, then we have
for the expressions of the entropy and enthalpy that correspond to
a universe which, if $H_0>\sqrt{4\pi GV_{SQ}}$, first expands in
an accelerated way with $w>-1$, then expands in a decelerating way
to finally progressively contract all the way down until it fades
out at an infinite time,
\begin{equation}
S_{SQ}(a_{-}, H_{0})=\frac{3H_0^2}{8\pi G\kappa}
-\frac{V_{SQ}}{\kappa}\ln\left[\left(\frac{a_{-}}{a_0}\right)^3\right]
,
\end{equation}
and again for this case
\begin{equation}
H_{SQ}(a_{-}, H_{0})=\frac{V_{SQ}}{\kappa}=k_B ,
\end{equation}
which is now positive definite.

Eq. (3.5) contains two different terms. The first term,
$S_{dS}=3H_0^2 k_B /(8\pi GV_{SQ})$, corresponds to a de Sitter
quantum entropy which diverges in the classical limit
$\hbar\rightarrow 0$. The second one is the same as the
statistical-mechanic entropy in Eq. (3.2) but with the sign
reversed. It would be worth comparing the first entropy term with
the Hawking formula for de Sitter space-time which is given by the
horizon area in Planck units, $S_{H}\propto
H_0^{-2}k_B/(\ell_P^2)$ [16]. At first sight the entropy term
$S_{dS}$ appears to be proportional to just the inverse of the
Hawking's formula. However, one can re-write $S_{dS}$ as
$S_{dS}=k_B/(2GH_0\bar{V}_{SQ})$, where $\bar{V}_{SQ}=V_{SQ}{\rm
V}_{dS}$, with ${\rm V}_{dS}$ the equivalent volume occupied by de
Sitter space-time with horizon at $r=H_0^{-1}$. Now,
$\bar{V}_{SQ}$ is the amount of sub-quantum energy contained in
that equivalent de Sitter volume, so that we must have
$\bar{V}_{SQ}=\hbar H_0$. It follows that $S_{dS}$ actually
becomes given by the horizon area in Planck units, too. It is
worth noticing that the temperature $T_{SQ}(a_{-}, H_{0})$ can
similarly be decomposed into two parts one of which is given by
the Gibbons-Hawking expression [16] $\hbar H_0/k_B$, and the other
corresponds to the negative volume deficit that the factor
$\exp(-2\pi GV_{SQ}t^2)$ introduces in the de Sitter space-time
volume.

We note that also for this kind of solution a universe with
$T_{SQ}(a_{-}, H_{0})=V_{SQ}a_0^3/k_B$ and
$S_{SQ}(a_{-},H_{0})=S_{dS}$ is left when we set $t=0$. If we let
$\hbar\rightarrow 0$, then $T_{SQ}(a_{-}, H_{0})\rightarrow 0$ and
$S_{SQ}(a_{-},H_{0})\rightarrow\infty$. On the other hand, it
follows from Eq. (3.5) that, as the universe evolves from the
initial size $a_0$, the initially positive entropy
$S_{SQ}(a_{-},H_{0})$ progressively decreases until it vanishes at
a time $t=t_* =H_0/(4\pi GV_{SQ})$, after which the entropy
becomes negative. This would mean a violation of the second law of
thermodynamics even on the current evolution of the universe which
is induced by quantum effects. Therefore the model that
corresponds to Eqs. (3.5) and (3.6) appears to be prevented by the
second law.

Finally, we consider the remaining solution $a_{+}=a_0\exp(H_0
t+2\pi GV_{SQ}t^2)$ which predicts a universe expanding in a
super-accelerated fashion all the time up to infinity with $w<-1$.
In this case we obtain
\begin{equation}
S_{SQ}(a_{+},H_{0})=\frac{3H_0^2}{8\pi G\kappa}
+\frac{V_{SQ}}{\kappa}\ln\left[\left(\frac{a_{+}}{a_0}\right)^3\right]
,
\end{equation}
with $3H_0^2/(8\pi G\kappa)=3H_0^2 k_B/(8\pi GV_{SQ})\propto S_H$,
and
\begin{equation}
H_{SQ}(a_{+},H_{0})=-\frac{V_{SQ}}{\kappa}=-k_B .
\end{equation}
All the above discussion on the relation of the sub-quantum
thermodynamical functions with the Hawking temperature and entropy
holds also in this case, with the sole difference that now
$S_{SQ}(a_{+},H_{0})$ and $T_{SQ}(a_{+},H_{0})$ are larger than
their corresponding Hawking counterparts. Again for this solution
a universe with $T_{SQ}(a_{+},H_{0})=\kappa a_0^3$ and
$S_{SQ}(a_{+},H_{0})=S_{dS}$ is left when we set $t=0$ whereas
$T_{SQ}(a_{+},H_{0})\rightarrow 0$ and
$S_{SQ}(a_{+},H_{0})\rightarrow\infty$ in the classical limit
$\hbar\rightarrow 0$. Moreover, such as it happens when $H_0 =0$,
there is here no violation of the second law for
$S_{SQ}(a_{+},H_{0})$, but $H_{SQ}(a_{+},H_{0})$ is again a
negative constant interpretable like a negative entropy that would
mark the onset of existing structures in the universe which are
capable to store and process information [15].

In any case, we have shown that the thermodynamical laws derived
in this article appear to preclude any model with $w>-1$ and so
leave only a kind of phantom universe with $w<-1$ as the only
possible cosmological alternative compatible with such laws. That
kind of model does not show however the sort of shortcomings,
including instabilities, negative kinetic field terms or the
future singularities named big rips, that the usual phantom models
have [17]. Since we have dealt with an essentially quantum system,
the violation of the dominant energy condition that leads to the
negative values of the enthalpy $H_{SQ}$ in the
thermodynamically-allowed models appears to be a rather benign
problem from which one could even get some interpretational
advantages. In fact, from Eqs. (2.9) - (2.11) we notice that the
violation of the dominant energy condition (DEC)
\begin{equation}
\rho+p=-V_{SQ} ,
\end{equation}
has an essentially quantum nature, so that such a violation
vanishes in the classical limit where $\hbar\rightarrow 0$. In
fact, it is currently believed that, even though classical general
relativity cannot be accommodated to a violation of the dominant
energy condition [18], such a violation can be admitted quantum
mechanically, at least temporarily. Moreover, since the violating
term $-V_{SQ}$ is directly related to the negentropy
$H_{SQ}=-k_B$, it is really tempting to establish a link between
that violation and the emergence of life in the universe. After
all, one cannot forget that if living beings are fed on with
negative entropy [15] then we ought to initially have some amount
of negentropy to make the very emergence of life a more natural
process which by itself satisfies the second law.

\subsection{Violation of classical DEC}

Thus, the quantum violation of the dominant energy condition has
not any classical counterpart and therefore is physically
allowable. We shall investigate in what follows the sense in which
that violation would permit the formation of Lorentzian wormholes.
Choosing the simplest mixed energy-momentum tensor components and
the ansatz that correspond to a static, spherically-symmetric
wormhole spacetime with vanishing shift function, $ds^2 = -dt^2
+e^{\lambda}dr^2 +r^2 d\Omega_2^2$ (where $d\Omega_2^2$ is the
metric on the unit two-sphere), we can obtain a wormhole spacetime
solution from the corresponding Einstein equations containing the
extra sub-quantum energy density and pressure, that is
\[-\frac{\lambda '}{r}e^{-\lambda} -\frac{1}{r^2}\left(e^{-\lambda} -1\right)=-\frac{8\pi
G}{3}\left(\frac{9r_0^2}{8\pi Gr^4}+\rho\right) \]
\[-\frac{1}{r^2}\left(e^{-\lambda} -1\right)=\frac{8\pi
G}{3}\left(\frac{3r_0^2}{8\pi Gr^4}+p\right) \]
\[-\frac{1}{2}e^{-\lambda}\frac{\lambda '}{r}=\frac{8\pi G}{3}\left(\frac{3r_0^2}{8\pi
Gr^4}+p\right) ,\] supplemented by the condition $\rho+p=-V_{SQ}$,
to obtain
\begin{equation}
ds^2 = -dt^2 +\frac{dr^2}{1-\frac{r_0^2}{r^2}+\ell_P^2 V_{SQ}r^2}
+r^2 d\Omega_2^2 ,
\end{equation}
with $r_0$ the radius of the spherical wormhole throat and
$\ell_P$ the Planck length. Note that if $\rho+p$ was positive
then no cosmic wormhole could be obtained, such as it happens for
the de Sitter space. Metric (3.10) is by itself nevertheless an
actual cosmic wormhole because, if that metric is written as
\begin{equation}
ds^2 =-dt^2 +d\ell^2 +r^2 d\Omega_2^2 ,
\end{equation}
then the new parameter [19]
\begin{equation}
\ell=\pm\int_{r_0}^{r}\frac{r' dr'}{\sqrt{r'^2 -r_0^2 +\ell_P^2
V_{SQ}r'^4}}
=\pm\frac{1}{2\ell_P\sqrt{V_{SQ}}}ln\left(\frac{2\ell_P
\sqrt{V_{SQ}}\sqrt{r^2-r_0^2+\ell_P^2 V_{SQ}r^4}+2\ell_P^2
V_{SQ}r^2 +1}{1+4\ell_P^2 V_{SQ}r_0^2}\right)
\end{equation}
goes from $-\infty$ (when $r=+\infty$) to zero (at $r=r_0$) and
finally to $+\infty$ (when $r=\infty$ again), such as it is
expected for a wormhole with a throat at $r=r_0$ which is
traversable and can be converted into a time machine. It can be
readily checked that for $\rho+p>0$ there is no metric like (3.12)
which can show these properties.

\subsection{Holographic models}
Holographic models which are related with the entropy of a dark
energy universe have been extensively considered [20,21]. We shall
discuss now the main equation that would govern the holographic
model for the quantum cosmic scenario. If we try to adjust that
model to the Li's holographic description for dark energy [20],
then we had to define the holographic sub-quantum model by the
relation
\begin{equation}
H^2= \frac{8\pi G\rho}{3}=4\pi GV_{SQ}\mu(t)^2
\ln\left(8GV_{SQ}R_h^2\right) ,
\end{equation}
where the future event horizon $R_h=a(t)\int_t^{\infty}dt'/a(t')$
is given by
\begin{equation}
R_h=\frac{e^{x^2}}{\sqrt{8GV_{SQ}}}\left[1-\Phi(x)\right],
\end{equation}
with $\Phi(x)$ the probability integral [22],
\begin{equation}
x=\frac{H_0}{\sqrt{8\pi GV_{SQ}}}+\sqrt{2\pi GV_{SQ}}t,
\end{equation}
and
\begin{equation}
\mu(t)^2=\frac{1}{1+3(1+w(t))\ln\left[1-\Phi\left(-\frac{1}{1+w(t)}\right)\right]}
.
\end{equation}
Note that: (1) $R_h\rightarrow\infty$ as $t\rightarrow\infty$ or
$V_{SQ}\rightarrow 0$, (2) in the latter limit $H^2\rightarrow 0$,
(3) $\mu(t)^2$ is no longer a constant because we are dealing with
a tracking model where the parameter $w$ depends on time, and (4)
the holographic model has no the problems posed by the usual
holographic phantom energy models. However, this formulation does
not satisfy the general holographic equation originally introduced
by Li which reads [20] $\rho\propto H^2\propto c^2/R^2$ (where $R$
is the proper radius of the holographic surface and $c$ is a
parameter of order unity that depends on $w$ according to the
relation $w=-(1+2/c)/3$), and therefore seems not satisfactory
enough. A better and quite simpler holographic description which
comes from saturating the original bound on entropy [23] and
conforms the general holographic equation stems directly from the
very definitions of the energy density (2.9) and the entropy
(3.7). Such a definition would read
\begin{equation}
\rho=\kappa S_{SQ}(a_+,H_0)=\frac{3H^2}{8\pi G}=\frac{3}{8\pi
GR_H^2} .
\end{equation}
It appears that if the last equality in Eq. (3.18) holds then the
holographic screen is related to the Hubble horizon rather than
the future event horizon or particle horizon. In order to confirm
that identification we derive now the vacuum metric that can be
associated to our ever-accelerating cosmic quantum model with the
ansatz $ds^2= -e^{\nu}dt^2 +e^{\lambda}dr^2+r^2 d\Omega_2^2$. For
an equation of state $p=w\rho$ the Einstein equations then are
\begin{equation}
e^{-\lambda}\left(\frac{\lambda '}{r}-\frac{1}{r^2}\right) +
\frac{1}{2r^2} =8\pi G\rho
\end{equation}
\begin{equation}
e^{-\lambda}\left(\frac{\nu '}{r}+\frac{1}{r^2}\right) -
\frac{1}{2r^2} =8\pi Gw\rho .
\end{equation}
We get finally the non-static metric
\begin{equation}
ds^2 =-\left(1-H^2 r^2\right)^{-(1+3w)/2}dt^2 + \frac{dr^2}{1-H^2
r^2} + r^2 d\Omega_2^2 ,
\end{equation}
which consistently reduces to the de Sitter static metric for
$w=-1$. It follows that there exists a time-dependent apparent
horizon at $r=H^{-1}$ playing in fact the role of a Hubble
horizon, like in the de Sitter case.

This holographic model has several advantages over the previous Li
model [20] and other models [21], including its: naturalness (it
has been many times stressed that choosing the Hubble horizon is
quite more natural than using, for the sake of mathematical
consistency, particle or future event horizons), simplicity (no ad
hoc assumption has been made), implication of an IR cutoff
depending on time, formal equivalence with Barrow's hyper
inflationary model [24] (but here respecting the thermodynamical
second law as, in this case, $S_{SQ}(a_+,H_0)$ increases with
time), and allowance of a unification between the present model
and that for dark energy from vacuum entanglement [25].

\subsection{Quantum cosmic models and entanglement entropy}

The latter property deserves some further comments. In fact, if we
interpret $a^3 V_{SQ}$ as the total entanglement energy of the
universe, due to the additiviness of the entanglement entropy, one
can then add up [25] the contributions from all existing
individual fields in the observable universe, so that the entropy
of entanglement $S_{{\rm Ent}}=\beta R_H^2$ (see comment after Eq.
(3.8)), with $\beta$ a constant including the spin degrees of
freedom of quantum fields in the observable volume of radius $R_H$
and a numerical constant of order unity. On the other hand, the
presence of a boundary at the horizon leads us to infer that the
entanglement energy ought to be proportional to the radius of the
associated spherical volume, i.e. $E_{{\rm Ent}}=\alpha R_H$ [25],
with $\alpha$ a given constant. We have then,
\begin{equation}
E_{{\rm Ent}}=a^3 V_{SQ}=\alpha R_H
\end{equation}
\begin{equation}
S_{{\rm Ent}}=\beta R_H^2 .
\end{equation}
It is worth noticing that one can then interpret the used
temperature as the entanglement temperature, so that $E_{{\rm
Ent}}=k_B T_{}(a_+)$. Now, integrating over $R_H$ the expression
for $dE_{{\rm Ent}}$ derived by Lee, Lee and Kim [25] from the
saturated black hole energy bound [26],
\begin{equation}
dE_{{\rm Ent}}=T_{{\rm Ent}}dS_{{\rm Ent}}
\end{equation}
(where $T_{{\rm Ent}}=(2\pi R_H)^{-1}$ is the Gibbons-Hawking
temperature), we consistently recover expression (3.22) for
$\alpha=\beta/\pi$. This result is also consistent with the
holographic expression introduced before. It follows therefore
that the quantum cosmic holographic model considered in the
present paper can be consistently interpreted as an entangled dark
energy holographic model, similar to the one discussed in Refs.
[25], with the sub-quantum potential $V_{SQ}$ playing the role of
the entanglement energy density.

Before closing up this section, it would be worth mentioning that
the recent data [27] seem to point to a value $w<-1$, with
$\dot{w}$ small and positive, just the result predicted in the
present letter. We in fact note that from Eq. (2.7) we obtain that
$\dot{w}=4\dot{H}^2/(3H^3) \propto t^{-3}$, at sufficiently large
time.

\section{conclusions and comments}
This paper deals with two new four-dimensional cosmological models
describing an accelerating universe in the spatially flat case.
The ingredients used for constructing these solutions are minimal
as they only specify a cosmic relativistic field described by just
Hilbert-Einstein gravity and the probabilistic quantum effects
associated with particles in the universe. While one of the models
is ruled out on general thermodynamical grounds as being
unphysical, the other model corresponds to an equation of state
$p=w\rho$ with parameter $w<-1$ for its entire evolution; that is
to say, this solution is associated with the so-called phantom
sector, showing however a future evolution of the universe which
is free from most of the problems confronted by usual phantom
scenarios; namely, violent instabilities, future singularities and
classical violations of energy conditions. We have shown
furthermore that the considered phantom model implies a more
consistent cosmic holographic description and the equivalence
between the discussed models and the entangled dark energy model
of the universe. Therefore we name our phantom model a {\it
benigner phantom} model.

Indeed, if the ultimate cause for the current speeding-up of the
universe is quantum entanglement associated with its matter and
radiation contents, then one would expect that the very existence
of the current universe implied violation of the Bell's
inequalities and hence the quantum probabilistic description
related to the sub-quantum potential considered in this work, or
the collapse of the superposed cosmic quantum state into the
universe we are able to observe, or its associated complementarity
between cosmological and microscopic laws, any other aspects that
may characterize a quantum system. The current dominance of
quantum repulsion over attractive gravity started at a given
coincidence time would then mark the onset of a new {\it quantum}
region along the cosmic evolution, other than that prevailed at
the big bang and early primeval universe, this time referring to
the quite macroscopic, apparently classical, large universe which
we live in. Thus, quite the contrary to what is usually believed,
quantum physics does not just govern the microscopic aspects of
nature but also the most macroscopic domain of it in such a way
that we can say that current life is forming part and is a
consequence of a true quantum system.

Observational data are being accumulated that each time more
accurately point to an equation of state for the current universe
which corresponds to a parameter whose value is very close to that
of the case of a cosmological constant, but still being less than
-1 [27]. It appears that one of the models considered in this
paper would adjust perfectly to such a requirement, while it does
not show any of the shortcomings that the customary phantom or
modified-gravity scenarios now at hand actually have. Therefore,
one is tempted to call for more developments to be made on such
benigner cosmological model, aiming at trying to construct a final
scenario which would consistently describe the current universe
and could presumably shed some light on what really happened
during the primordial inflationary period as well.

\acknowledgements

\noindent This work was supported by MEC under Research Project
No. FIS2005-01181. Alberto Rozas-Fern\'{a}ndez acknowledges
support from MEC FPU grant No. AP2004-6979. The authors benefited
from discussions with C. Sig\"{u}enza and G. Readman.

\pagebreak

\renewcommand{\theequation}{B-\arabic{equation}}
  % redefine the command that creates the equation no.
  \setcounter{equation}{0}  % reset counter
  \section*{APPENDIX}  % use *-form to suppress numbering

%\pagebreak
%\begin{center}
%\appendix{\bf APPENDIX: ON THE BACKGROUND THEORIES}
%\end{center}

In this Appendix we shall consider new fundamental aspects that
strengthen the consistency and provide further physical motivation
to the general model reviewed in Sec. II. These new aspects
concern both the use of a sub-quantum potential model derived from
the application of the Klein-Gordon equation, and the background
relativistic theory associated with the cosmic quantum models.

\subsection{The Klein-Gordon sub-quantum model}
We note here that, although for some time in the past it was
generally believed that the Klein-Gordon equation was unobtainable
from the Bohm formalism [28], in recent years the Klein-Gordon
equation has found satisfactory causal formulations. The solution
presented in [29] by Horton \emph{et al}. has to introduce the
causal description of time-like flows in an Einstein-Riemann space
(otherwise the probability current can assume negative values of
its zeroth component and is not generally time-like). However,
there exists a causal Klein-Gordon theory in Minkowski space [30]
where this is achieved by introducing a cosmological constant as
an additional assumption which is justified in view of recent
observations. Therefore, it makes perfect sense to use a
Klein-Gordon equation in our model [3]. Moreover, the nonclassical
character of the current whose continuity equation is derived from
the purely imaginary part of the expression resulting from the
application of the Klein-Gordon equation to the wave function is
guaranteed by the fact that one can never obtain the classical
limit by making $\hbar\rightarrow 0$. Thus, no classical verdict
concerning that current of the kind pointed out by Holland [28]
can be established. On the other hand, having a material object
whose trajectory escapes out the light cone [28] cannot be used as
an argument in favour of the physical unacceptability of the
model. Quite the contrary, it expresses its actual essentially
quantum content, much as the quite fashionable entangled states of
sharp quantum theory seemed at first sight violate special
relativity and then turned out to be universally accepted. In both
cases, physics is preserved because we are not dealing with real
signaling. Actually, in Sec. III we have shown that our cosmic
models can be also interpreted as being originated from the
entanglement energy of the whole universe, without invoking any
other cause.

\subsection{Quantum theory of special relativity}
Consistent tachyonic theories for dark energy are grounded on
special theory of relativity in such a way that all the physics
involved at them stems from Einstein relativity. Our cosmic
quantum models actually come from a generalization from tachyonic
theories for which the corresponding background relativistic
description ought to contain the quantum probabilistic footprint.
Thus, in order to check their consistency, viability and properly
motivate the models reviewed in Sec. II, one should investigate
the characteristics of the quantum relativistic theory on which
they are based. In what follows we shall consider in some detail
the basic foundations of that background quantum relativity.

Actually, there are two ways for defining the action of a free
system endowed with a rest mass $m_0$ [31]. The first one is by
using the integral expression for the Lagrangian $L =\int pdv$,
with the momentum $p$ derived from the Hamilton-Jacobi equation,
and inserting it in the expression $S=\int_{t_1}^{t_2}Ldt$. The
second procedure stems from the definition $S=\beta\int_a ^b ds$,
where $ds$ is the line element and the proportionality constant
$\beta=m_0 c$ is obtained by going to the non-relativistic limit.
The strategy that we have followed here is to apply the first
procedure to derive an integral expression for $S$ in the case of
a Hamilton-Jacobi equation containing an extra quantum term and
then obtain the expression for $ds$ by comparing the resulting
expression for $S$ with that is given by the second procedure.

As mentioned above, a Hamilton-Jacobi equation with the quantum
extra term can be obtained by applying the Klein-Gordon equation
to a quasiclassical wave function $\Psi=R(r,t)\exp(iS(r,t)/\hbar)$
[32], where $R(r,t)$ is the quantum probability amplitude and
$S(r,t)$ is the classical action. By the second of the above
procedures and $L_Q =-m_0 c^2 E(\varphi,k)$, we immediately get
for the general spacetime metric
\begin{equation}
ds=E(\phi,k)dt,
\end{equation}
which consistently reduces to the metric of special relativity in
the limit $\hbar\rightarrow 0$. If we take the above line element
as invariant, then we obtain for time dilation
\begin{equation}
dt=\frac{E(k)dt_0}{E(\varphi,k)} ,
\end{equation}
in which $E(k)$ is the complete elliptic integral of the second
kind [22].

A key question that arises now is, does the quantum relativistic
description and hence our cosmic quantum models satisfy Lorentz
invariance? What should be invariant in the present case is the
quantity
\begin{equation}
I=ctE\left(\arcsin\sqrt{\frac{c^2 t^2 -x^2}{c^2 t^2}}, k\right)
\end{equation}
If we would choose a given transformation group in terms of
hyperbolic or elliptic functions which leaves invariant (such as
it happens for Lorentz transformations) the usual relativistic
combination $c^2 t^2 -x^2 = c^2 t'^2 - x'^2$, then we obtained
\begin{equation}
I=cQ(t',x',)E\left(\arcsin\frac{\sqrt{c^2 t'^2 -x'^2}}{cQ(t',x')}
, k\right) ,
\end{equation}
where $Q(t',x')\equiv Q(t',x',\Psi)$ is the expression for the
transformation of time $t$ in terms of hyperbolic or elliptic
functions. It would follow
\begin{equation}
\left(\frac{I}{cQ(t',x')}\right)^{-1} =\frac{\sqrt{c^2 t'^2
-x'^2}}{cQ(t',x')} ,
\end{equation}
with $\left(\right)^{-1}$ denoting the inverted function
associated to the elliptic integral of the second kind, generally
one of the Jacobian elliptic functions or a given combination of
them [22]. Thus, the quantity $I$ can only be invariant under the
chosen kind of transformations in the classical limit where $k=1$.
Therefore, a quantum relativity built up in this way would clearly
violate Lorentz invariance, at least if we take usual classical
values for the coordinates.

In order to obtain the wanted transformation equations we first
notice that if we take the coordinate transformation formulas in
terms of the usual hyperbolic or some elliptic functions of the
rotation angle $\Phi$ one can always re-express the invariant
quantity $I$ of Einstein special relativity in the form
\begin{equation}
I=cQ(t',x')E\left(\arcsin\left(\frac{\sqrt{c^2
t'^2-x'^2}}{cQ(t',x')}\right)^{-1} , k\right) .
\end{equation}
>From Eq. (B-6) one can write

\[\left(\frac{I}{cQ(t',x')}\right)^{-1} =\left(\frac{\sqrt{c^2
t'^2-x'^2}}{cQ(t',x')}\right)^{-1}\] and hence
\begin{equation}I=\sqrt{c^2
t'^2-x'^2}=ct'E\left(\arcsin\left(\frac{\sqrt{c^2
t'^2-x'^2}}{ct'}\right)^{-1} , k\right) ,
\end{equation}
that is $I$ would in fact have the form of the Einstein
relativistic invariant. If we interpret the coordinates entering
Eq. (B-7) as quantum-mechanical coordinates, then our quantum
expression for the invariant $I$ given by Eq. (B-3) can be
directly obtained from the last equality by making the replacement
\begin{equation}
\sqrt{1-\frac{x^2}{c^2
t^2}}=E\left(\arcsin\sqrt{1-\frac{x_{clas}^2}{c^2 t_{clas}^2}} ,
k\right)
\end{equation}
or
\begin{equation}
\left(\sqrt{1-\frac{x^2}{c^2
t^2}}\right)^{-1}=\sqrt{1-\frac{x_{clas}^2}{c^2 t_{clas}^2}} ,
\end{equation}
where The notation $( )^{-1}$ again means inverted function of the
elliptic integral of the second kind, and if the coordinates
entering the right-hand-side are taken to be classical
coordinates, then those on the left-hand-side must still in fact
be considered to be quantum-mechanical coordinates. Classical
coordinates are those coordinates used in Einstein special
relativity and set the occurrence of a classical physical event in
that theory. By quantum coordinates we mean those coordinates
which are subject to quantum probabilistic uncertainties and would
define what one may call a quantum physical event: i.e. that event
which is quantum-mechanically spread throughout the whose existing
spacetime with a given probability distribution fixed by the
boundaries specifying the extent and physical content of the
system.

In what follows we will always express all equations in terms of
classical coordinates and therefore, for the sake of simplicity,
we shall omit the subscript "clas" from them. The equivalence
relation given by expressions (B-8) and (B-9) is equally valid for
primed and non primed coordinates and should be ultimately related
with the feature that for a given, unique time, $t$ or $t'$, the
position coordinate, $x$ or $x'$, must be quantum-mechanically
uncertain.
>From the equalities (B-8) and (B-9) for primed coordinates we get
then an expression for $I'$ in terms of classical coordinates
\begin{equation}
I'=ct'E\left(\arcsin\frac{\sqrt{c^2 t'^2-x'^2}}{ct'} , k\right) ,
\end{equation}
which shows the required invariance and in fact becomes the known
relativistic result $I'=\sqrt{c^2 t'^2-x'^2}$ in the classical
limit $\hbar\rightarrow 0$.

>From expressions (B-8) and (B-9) we also have
\begin{equation}
1-\frac{V^2}{c^2}=E(\varphi,k)^2\rightarrow
\frac{V}{c}=\sqrt{1-E(\varphi,k)^2}=\tanh\Phi ,
\end{equation}
where $V$ is velocity, $\varphi=\arcsin\sqrt{1-\frac{x^2}{c^2
t^2}}$ and we have specialized to using the usual hyperbolic
functions. Whence $\cosh\Phi=1/E(\varphi,k)$,
$\sinh\Phi=\sqrt{1-E(\varphi,k)^2}/E(\varphi,k)$, and from the
customary hyperbolic transformation formulas for coordinates
\begin{equation}
x=x'\cosh\Phi+ct'\sinh\Phi, \;\; ct=ct'\cosh\Phi+x'\sinh\Phi ,
\end{equation}
we derive the new quantum relativistic transformation equations
\begin{equation}
x=\frac{x'+ct'\sqrt{1-E(\varphi,k)^2}}{E(\varphi,k)} ,\;\;
ct=\frac{ct'+x'\sqrt{1-E(\varphi,k)^2}}{E(\varphi,k)} .
\end{equation}

Had we started with formulas expressed in terms of the Jacobian
elliptic functions [22], such that:
\begin{equation}
\frac{V}{c}={\rm sn}(\Phi,k)=\sqrt{1-E(\varphi,k)^2}
\end{equation}
\begin{equation}
x=x'{\rm nc}(\Phi,k)+ct'{\rm sc}(\Phi,k), \;\; ct=ct'{\rm
nc}(\Phi,k)+x'{\rm sc}(\Phi,k) ,
\end{equation}
then we had again obtained Eqs. (B-13), so confirming the
quantum-mechanical character of the coordinates entering the
left-hand-side of Eqs. (B-8) and (B-9). The above derived
expressions are not yet the wanted expressions as they still
contain an unnecessary element of classicality due to the feature
that when using quantum-mechanical coordinates for the derivation
of the velocity $V$ setting $x=0$ the unity of the left-hand-side
of Eq. (B-8) would correspond to the complete elliptic integral of
the second kind $E(k)$ [22]. Thus, we finally get for the
transformation equations
\begin{eqnarray}
&&x=\frac{\left(x'+ct'\sqrt{1-E(\varphi,k)^2}\right)E(k)}{E(\varphi,k)}\nonumber\\
&&ct=\frac{\left(ct'+x'\sqrt{1-E(\varphi,k)^2}\right)E(k)}{E(\varphi,k)}
,
\end{eqnarray}
that are the wanted final expressions in terms of classical
coordinates which in fact reduce to the known Lorentz
transformations in the classical limit $\hbar\rightarrow 0$. From
the formula for time transformation we in fact get time dilation
to be the same as that (Eq. (B-2)) directly obtained from the
metric when referring to two events occurring at one and the same
point $x'$, i.e.
\begin{equation}
\Delta t=\frac{E(k)\Delta t_0}{E(\varphi,k)} ,
\end{equation}
and from that for space transformation the formula for length
contraction referred to one and the same time $t'$
\begin{equation}
\Delta\ell=\frac{E(\varphi,k)\Delta\ell_0}{E(k)} .
\end{equation}
In any case, the quantum effects would be expected to be very
small, that is usually $k$ is generally very close to unity for
sufficiently large rest masses of the particles.

For the sake of completeness we shall derive in what follows the
transformation of velocity components one can also derive from the
coordinate transformations (B-16) that, if space and time
themselves are subject to the quantum-mechanical uncertainties,
they should be now given as \[v_x =\frac{v'_x
+c\sqrt{1-E(\varphi,k)^2}}{1+\frac{v'_x}{c}\sqrt{1-E(\varphi,k)^2}}
\]
\begin{equation}
v_y =\frac{v'_y
E(\varphi,k)}{E(k)\left(1+\frac{v'_x}{c}\sqrt{1-E(\varphi,k)^2}\right)}
\end{equation}
\[v_z =\frac{v'_z
E(\varphi,k)}{E(k)\left(1+\frac{v'_x}{c}\sqrt{1-E(\varphi,k)^2}\right)}
,\] which reduce once again to the well-known velocity
transformation law of Einstein special relativity. Even though
they are quantitatively distinct of the latter transformation law,
Eqs. (B-19) behave qualitatively in a similar fashion and produce
the analogous general velocity addition law as in Einstein special
relativity.

We finally turn to the essentials of the relativistic mechanics
and find the formulas for momentum and energy that must be
satisfied by the cosmic quantum models to be given by
\begin{equation}
p=\frac{\partial L}{\partial v}=\frac{m_0
c\sqrt{1-k^2\left(1-\frac{v^2}{c^2}\right)}}{\sqrt{1-\frac{v^2}{c^2}}}
\end{equation}
\begin{eqnarray}
&&E=pv-L=\frac{m_0 c^2}{\sqrt{1-\frac{v^2}{c^2}}}\times\nonumber\\
&&\left[\frac{v}{c}\sqrt{1-k^2\left(1-\frac{v^2}{c^2}\right)} +
\sqrt{1-\frac{v^2}{c^2}}E(\varphi,k)\right] .
\end{eqnarray}

Obviously, these expressions reduce to $p=m_0 v/\sqrt{1-v^2/c^2}$
and $E=m_0 c^2/\sqrt{1-v^2/c^2}$, respectively, in the limit
$\hbar\rightarrow 0$. Moreover, if we set $v=0$ then $p=V_{Q}/c$
and $E=m_0 c^2 E(k)$ which become, respectively, $0$ and $m_0 c^2$
when $\hbar\rightarrow 0$. It follows then that our quantum
special relativistic model has the expected good limiting
behavior.

Unless for rather extreme cases the value of parameter $k$ is very
close to unity and therefore the corrections to the customary
expressions induced by the present model should be expected to be
very small locally. However, they could be perhaps detectable in
specially designed experiments using extremely light particles.

The main conclusion that can be drawn from the above discussion is
that whereas Lorentz invariance appears to be violated in our
quantum description if classical coordinates are considered, such
an invariance is preserved when one uses quantum coordinates in
that description.


\begin{references}
\bibitem {1} Edmund J. Copeland, M. Sami and Shinji Tsujikawa,
Int.J.Mod.Phys.D15 (2006) 1753.

\bibitem {2} A.H. Guth, Phys. Rev. D23 (1981) 347 .

\bibitem {3} P.F. Gonz\'{a}lez-D\'{\i}az, {\it Dark energy without dark
energy}, AIP Conf. Proc. 878 (2006) 227; P.F.
Gonz\'{a}lez-D\'{\i}az and Alberto Rozas-Fern\'{a}ndez, Phys.Lett.
B641 (2006) 134.

%\cite{Lima:2004wf}
\bibitem{Lima:2004wf}
  J.~A.~S.~Lima and J.~S.~Alcaniz,
  %``Thermodynamics and spectral distribution of dark energy,''
  Phys.\ Lett.\  B {\bf 600} (2004) 191.

%\cite{GonzalezDiaz:2004eu}
\bibitem{GonzalezDiaz:2004eu}
  P.~F.~Gonzalez-Diaz and C.~L.~Siguenza,
  %``Phantom thermodynamics,''
  Nucl.\ Phys.\  B {\bf 697} (2004) 363.
  %%CITATION = NUPHA,B697,363;%%

%\cite{Nojiri:2004pf}
\bibitem{Nojiri:2004pf}
  S.~Nojiri and S.~D.~Odintsov,
  %``The final state and thermodynamics of dark energy universe,''
  Phys.\ Rev.\  D {\bf 70} (2004) 103522 .

%\cite{Izquierdo:2005ku}
\bibitem{Izquierdo:2005ku}
  G.~Izquierdo and D.~Pavon,
  %``Dark Energy And The Generalized Second Law,''
  Phys.\ Lett.\  B {\bf 633} (2006) 420 .


%\cite{Izquierdo:2006mt}
\bibitem{Izquierdo:2006mt}
  G.~Izquierdo and D.~Pavon,
  %``The generalized second law in dark energy dominated universes,''
  arXiv:gr-qc/0612092.
  %%CITATION = GR-QC/0612092;%%

%\cite{Gong:2006ma}
\bibitem{Gong:2006ma}
  Y.~Gong, B.~Wang and A.~Wang,
  %``Thermodynamical properties of the universe with dark energy,''
  JCAP {\bf 0701} (2007) 024 .

%\cite{Gong:2006sn}
\bibitem{Gong:2006sn}
  Y.~Gong, B.~Wang and A.~Wang,
  %``On thermodynamical properties of dark energy,''
  Phys.\ Rev.\  D {\bf 75} (2007) 123516 .

\bibitem {11} P.F. Gonz\'{a}lez-D\'{\i}az, Phys. Rev. D69 (2004) 103512 .

\bibitem {12} J.S. Bagla, H.K. Jassal and T. Padmanabhan, Phys. Rev. D67, 063504 (2003).

\bibitem {13} T. Padmanabhan, Phys. Rev. D66 (2002) 021301; T. Padmanabhan and T.R. Choudhury,
Phys. Rev. D66 (2002) 081301 .

\bibitem{Caldwell:1999ew}
  R.~R.~Caldwell,
  Phys.\ Lett.\ B {\bf 545} (2002) 23;
  %%CITATION = ASTRO-PH 9908168;%%
  %%Cited 363 times in SPIRES-HEP
%\cite{Carroll:2003st}
%\bibitem{Carroll:2003st}
  S.~M.~Carroll, M.~Hoffman and M.~Trodden,
  %
  Phys.\ Rev.\ D {\bf 68} (2003) 023509 .
  %%CITATION = ASTRO-PH 0301273;%%
  %%Cited 218 times in SPIRES-HEP

\bibitem {15} E. Schr\"{o}dinger, {\it What is Life?. The Physical
Aspects of the Living Cell} (Cambridge University Press,
Cambridge, UK, 1959); S.A. Kauffman, in {\it "What is Life?". The
Next Fifty Years} (Cambridge University Press, Cambridge, UK,
1995); L. Smolin, {\it The LIfe of the Cosmos} (Phoenix, London,
UK, 1997).

%\cite{Gibbons:1977mu}
\bibitem{Gibbons:1977mu}
  G.~W.~Gibbons and S.~W.~Hawking,
  %``Cosmological Event Horizons, Thermodynamics, And Particle Creation,''
  Phys.\ Rev.\  D {\bf 15} (1977) 2738 .
  %%CITATION = PHRVA,D15,2738;%%

\bibitem {17} R.R. Caldwell, Phys. Lett. B545 (2002) 23 ; R.R.
Caldwell, M. Kamionkowski and N.N. Weinberg, Phys. Rev. Lett. 91
(2003) 071301; P.F. Gonz\'{a}lez-D\'{i}az, Phys. Lett. B586 (2004)
1; Phys. Rev. D69, 063522 (2004); S.M. Carroll, M. Hoffman and M.
Trodden, Phys. Rev. D68 (2003) 023509; S. Nojiri and S.D.
Odintsov, Phys. Rev. D70 (2004) 103522 .

%\cite{Hawking:1973uf}
\bibitem{Hawking:1973uf}
  S.~W.~Hawking and G.~F.~R.~Ellis,
  %``The Large scale structure of space-time,''
%\href{http://www.slac.stanford.edu/spires/find/hep/www?irn=6991262}{SPIRES entry}
{\it Cambridge University Press, Cambridge, UK, 1973}.

\bibitem {19} M.S. Morris and K.S. Thorne, Am. J. Phys. 56 (1988)
395 .

\bibitem {20} M. Li, Phys. Lett. B603 (2004) 1 .

\bibitem {21} D. Pav\'{o}n and W. Zimdahl, Phys. Lett. B628 (2005)
206; A.G. Cohen, D.B. Kaplan and A.E. Nelson, Phys. Rev. Lett. 82
(1999) 4971 .

\bibitem {22} M. Abramowitz and I.A. Stegun, {\it Handbook of
Mathematical Functions} (Dover, New York, USA, 1972); I. S.
Gradshteyn and I. M. Ryzhik, Tables and Integrals, Series and
Products (Academic Press, New York, 1980).


\bibitem {23} P.F. Gonz\'{a}lez-D\'{\i}az, Phys. Rev. D27 (1983) 3042 .

\bibitem {24} J.D. Barrow, Phys. Lett. B180 (1986) 335; B193
(1987) 285 .

\bibitem {25} J.-W. Lee, J. Lee and H.-Chan Kim, JCAP 0708 (2007)
005; {\it Quantum informational dark energy: Dark energy from
forgetting}, hep-th/0709.0047 .

\bibitem {26} S. Mukohyama, M. Seriu and H. Kodama, Phys. Rev. D55
(1997) 7666 .


\bibitem{Mortlock:2000zu}
  D.~J.~Mortlock and R.~L.~Webster,
   %``The statistics of wide-separation lensed quasars,''
  %
  Mon.\ Not.\ Roy.\ Astron.\ Soc.\  {\bf 319} (2000) 872   [arXiv:astro-ph/0008081];
  %%CITATION = ASTRO-PH 0008081;%%
  %%Cited 4 times in SPIRES-HEP

%\cite{Riess:1998cb}
%\bibitem{Riess:1998cb}
  A.~G.~Riess {\it et al.}  [Supernova Search Team Collaboration],
  % ``Observational Evidence from Supernovae for an Accelerating Universe and a Cosmological Constant,''%
  Astron.\ J.\  {\bf 116} (1998) 1009
  [arXiv:astro-ph/9805201];
  %%CITATION = ASTRO-PH 9805201;%%
  %%Cited 1219 times in SPIRES-HEP

%\cite{Perlmutter:1998np}
%\bibitem{Perlmutter:1998np}
  S.~Perlmutter {\it et al.}  [Supernova Cosmology Project Collaboration],
  % ``Measurements of Omega and Lambda from 42 High-Redshift Supernovae,''%
  Astrophys.\ J.\  {\bf 517} (1999) 565
  [arXiv:astro-ph/9812133];
  %%CITATION = ASTRO-PH 9812133;%%
  %%Cited 1444 times in SPIRES-HEP

%\cite{Tonry:2003zg}
%\bibitem{Tonry:2003zg}
  J.~L.~Tonry {\it et al.}  [Supernova Search Team Collaboration],
  % ``Cosmological Results from High-z Supernovae,''%
  Astrophys.\ J.\  {\bf 594} (2003) 1   ;
  %%CITATION = ASTRO-PH 0305008;%%
  %%Cited 299 times in SPIRES-HEP

%\cite{Spergel:2003cb}
%\bibitem{Spergel:2003cb}
  D. N.~Spergel {\it et al.}  [WMAP Collaboration],
   %``First Year Wilkinson Microwave Anisotropy Probe (WMAP) Observations: Determination of
   % Cosmological  Parameters,''
  Astrophys.\ J.\ Suppl.\  {\bf 148} (2003) 175;
  %%CITATION = ASTRO-PH 0302209;%%
  %%Cited 746 times in SPIRES-HEP

%\cite{Bennett:2003bz}
%\bibitem{Bennett:2003bz}
  C.~L.~Bennett {\it et al.},
   %``First Year Wilkinson Microwave Anisotropy Probe (WMAP) Observations:
   %Preliminary Maps and Basic Results,''
  %
  Astrophys.\ J.\ Suppl.\  {\bf 148} (2003) 1;
  %%CITATION = ASTRO-PH 0302207;%%
  %%Cited 1207 times in SPIRES-HEP

%\cite{Tegmark:2003ud}
%\bibitem{Tegmark:2003ud}
  M.~Tegmark {\it et al.}  [SDSS Collaboration],
   %``Cosmological parameters from SDSS and WMAP,''
  Phys.\ Rev.\ D {\bf 69}, 103501 (2004).
  %%CITATION = ASTRO-PH 0310723;%%
  %%Cited 514 times in SPIRES-HEP
%%%%%%



\bibitem {28} P. R. Holland, \emph{The quantum theory of motion} (CUP,
Cambridge, UK, 1993).

\bibitem {29} G. Horton, C. Dewdney, A. Nesteruk, J. Phys. A 33 (2000) 7337 .

\bibitem {30} G. Gr\"{o}ssing, Phys. Lett. A 296 (2002) 1 .

\bibitem {31} L.D. Landau and E.M. Lifshitz, {\it The classical
theory of fields} (pergamon Press, Oxford, UK, 1975) .

\bibitem {32} D.J. Bohm, {\it Quantum Theory} (Dover, Prentice
Hall, New York, USA, 1951); Phys. Rev. 85 (1952) 180 .


\end{references}
\end{document}